\documentclass[fleqn,usenatbib]{mnras}

\usepackage{newtxtext,newtxmath}
\usepackage[T1]{fontenc}
\usepackage{xcolor}
\usepackage{booktabs}
{}
\usepackage{cancel}
\usepackage{soul,xcolor}
\setstcolor{red}
\usepackage{hyperref}

\DeclareRobustCommand{\VAN}[3]{#2}
\let\VANthebibliography\thebibliography
\def\thebibliography{\DeclareRobustCommand{\VAN}[3]{##3}\VANthebibliography}

\usepackage{graphicx}	% Including figure files
\usepackage{amsmath}	% Advanced maths commands

\usepackage{siunitx}
\usepackage{threeparttable}
\usepackage{comment}

\title[B-fields in L328 neighborhood]{Magnetic fields in the close neighborhood of LDN\,328 core}

\author[Gupta et al.]{
Shivani Gupta,$^{1,2}$\thanks{E-mail: shivani.gupta@iiap.res.in}
Archana Soam,$^{1}$
Janik Karoly,$^{3,4}$
Chang Won Lee$^{5,6}$
\\
$^{1}$Indian Institute of Astrophysics, II Block, Koramangala, Bengaluru 560034, India \\
$^{2}$Pondicherry University, R.V. Nagar, Kalapet, 605014, Puducherry, India\\
$^{3}$Department of Physics and Astronomy, University College London, WC1E 6BT London, UK\\
$^{4}$Jeremiah Horrocks Institute, University of Central Lancashire, Preston PR1 2HE, UK\\
$^{5}$Korea Astronomy and Space Science Institute (KASI) 776 Daedeokdae-ro, Yuseong-gu, Daejeon 34055, Republic of Korea\\
$^{6}$University of Science and Technology, Korea (UST), 217 Gajeong-ro, Yuseong-gu, Daejeon 34113, Republic of Korea\\
}

\date{Accepted XXX. Received YYY; in original form ZZZ}

\pubyear{\the\year{}}

\begin{document}
\label{firstpage}
\pagerange{\pageref{firstpage}--\pageref{lastpage}}
\maketitle

% Abstract of the paper
\begin{abstract}
Linearly polarized dust emission traces the plane-of-sky magnetic ﬁeld structure, thus allowing us to investigate the role of magnetic fields in the formation and evolution of cloud cores. In this work, we present observations of dust polarization  at 850 $\mu$m wavelength toward the LDN\,328 (hereafter, L328) core and its neighborhood regions, classified as R1, R2, and R3, using the James Clerk Maxwell Telescope (JCMT) with SCUBA-2/POL-2. This study extends our previous work on magnetic fields in L328 across different spatial scales. We used the JCMT/POL-2 data in the near vicinity of the L328 core to study the magnetic fields in its nearby regions. 
% The L328 core seems to be connected to the neighboring clumps, which is evident by a bridge-like structure, and the magnetic field seem to follow the bridge. 
We identified clumps in these regions using the `FellWalker' algorithm and derived the dust temperature and column density by fitting the spectral energy distribution (SED) using combined JCMT and $Herschel$ dust continuum maps. We analyzed the magnetic field morphology in the vicinity of the L328 core and employed the structure function analysis to determine the magnetic field dispersion angle. We used the modified Davis-Chandrasekhar-Fermi (DCF) method to estimate magnetic field strengths and also derived the mass-to-flux ratio and Alfvén Mach number.
%The criticality of the cores is also determined using estimated field strength.
\end{abstract}


%%%%%%%%%%%%%%%%%%%%%%%%%%%%%%%%%%%%%%%%%%%%%%%%%%

%%%%%%%%%%%%%%%%% BODY OF PAPER %%%%%%%%%%%%%%%%%%

\section{Introduction}

In the current understanding of star formation, clumps within molecular clouds serve as the precursors to dense cores and, ultimately, to formed stars {\citep{2003ARA&A..41...57L}}. These clumps, typically larger structures with hydrogen volume densities of n(H$_2$) > 10$^2$-10$^4$ cm$^{-3}$, are expected to fragment into smaller, denser cores as they evolve {\citep{2013MNRAS.432.3495V}}. This fragmentation is influenced by a combination of turbulence, gravity, thermal pressure, and magnetic fields (B-fields), determining whether these clumps will efficiently collapse into star-forming cores {\citep{2008ApJ...684..395H,2019FrASS...6....5H,2023ASPC..534..193P}}. However, not all clumps necessarily evolve into gravitationally bound cores or active star-forming regions. Some remain quiescent, while others may disperse before forming stars. Understanding the physical conditions, stability, and fragmentation properties of clumps is therefore crucial to constraining the initial conditions of star formation \citep{2003MNRAS.339..577B}. While the relative importance of B-fields versus turbulence remains debated, their role in clump evolution and star formation efficiency is critical {\citep{2024JApA...45...17S}}.

Sub-mm observations are an essential tool for tracing high-extinction (${A_{\rm V}}$) regions, as cooler dust grains emit thermal radiation, allowing us to trace the denser parts of the cloud, such as clumps or cores.  Since clumps are primarily composed of dust and gas, their structures are best studied at sub-mm wavelengths, where their B-fields can be analyzed through polarization measurements.  The plane-of-the-sky B-field ($B_{\rm{pos}}$) is measured by linear polarization of light by aligned dust grains \citep{1949Sci...109..166H, 1949ApJ...109..471H}. So far, the Radiative Torque Alignment (RAT) is the most accepted mechanism explaining the dust grain alignment in the ISM \citep{1976Ap&SS..43..291D, 2007MNRAS.378..910L,2008MNRAS.388..117H, ALV2015}. The minor axes of the dust grains align with the ambient B-field. At longer wavelengths (far-infrared to mm), the B-field orientation is inferred by rotating the polarization angle by 90$^\circ$ as thermal dust emission is polarized along the grains’ major axes that are perpendicular to the B-field.

In this work, we investigate the morphology and strength of the B-field in the nearest neighborhood of the L328 core using the JCMT 850 $\mu$m dust polarisation data.  Two prestellar cores and a low-mass star are found to be forming within the L328 core \citep [see] [for more details]{lee2018high}.
Earlier studies have analysed the B-field morphology in L328 using ${Planck}$, optical, NIR, and sub-mm (JCMT) polarisation data \citep{2015A&A...573A..34S, soam2015first,  2016A&A...594A...1P, 2016A&A...594A..26P,kumar2023magnetic,2025MNRAS.539.3493G}. The connection of B-fields across different wavelengths and spatial scales is illustrated in \cite{2025MNRAS.539.3493G}. To understand the surrounding regions of this core, we identified the dense, clumpy structures and estimated their temperature, column density, and volume density. Further, we analyzed the B-field structures in these clumps using our sub-mm polarisation measurements.

The structure of the paper is as follows: in Section \ref{sec2}, we present the observations and data reduction. Section \ref{sec3}, elaborates the analysis, results, and discussion. Finally, we summarize our results in Section \ref{summary}.

\section{Observations and data reduction}\label{sec2}

The observations at 850 $\mu$m were conducted with SCUBA-2/POL-2 on the JCMT in 2018 March (M18AP033; PI: Archana Soam) and in 2019 May and June (M19AP014; PI: Chang Won Lee). The weather conditions during observations were split between $\tau_{225}$<0.05 and 0.05<$\tau_{225}$ <0.08, where $\tau_{225}$ is the atmospheric opacity at 225 GHz. The data were collected using the Daisy-map mode  \citep{holland2013}, specifically optimized for POL-2 observations \citep{friberg2016}. The total integration time for a single ﬁeld was $\sim$ 31 minutes, with a total of 17 repeats, resulting in an on-source integration time of $\approx$8.8~hrs. This work focuses exclusively on the 850 $\mu$m data. Further details of the observations and data reduction can be found in \citet{2025MNRAS.539.3493G}.

A polarization vector catalog was created from the final Stokes \textit{I}, \textit{Q}, and \textit{U} maps by binning up from the 4\arcsec\,pixel size to 12\arcsec, which approximates the beam size 14\farcs1. This process reduces the number of vectors by combining vectors within each 12\arcsec\,pixel and also decreasing the noise level. Specifically, for plotting, we selected vectors with an intensity-to-error ratio ($I/\delta I$) > 10 and a polarization-to-error ratio {2 < ($P/\delta P$) < 25}.

In order to convert the native map units of pW to astronomical units, a flux calibration factor (FCF) of 495 Jy/beam/pW was used \citep{mairs2021}, multiplied by a factor of 1.35 to account for POL-2 being inserted into the beam. The peak values of total and polarized intensities are found to be $\sim$100 mJy beam$^{-1}$ and $\sim$11 mJy beam$^{-1}$, respectively. The rms noise of the background region in the Stokes \textit{I}, \textit{Q}, and \textit{U} maps is measured to be $\sim$6.26 mJy/beam, 5.27 mJy/beam, and 5.75 mJy/beam, respectively. These values were determined by selecting a region about 1\arcmin\,from the center of each corresponding map, where the signal remained relatively constant.

The values for the debiased degree of polarization $P$ were calculated using the modified asymptotic estimator \citep{plaszcz2014,montier2015}
\begin{equation}
P = \frac{1}{I} [PI - 0.5\sigma^{2}(1-e^{-(PI/\sigma)^{2}}/PI ] \,\, ,
\end{equation}
\noindent
where \textit{I}, \textit{Q}, and \textit{U} are the Stokes parameters, and $\sigma^{2}=(Q^{2}\sigma_{Q}^{2}+U^{2}\sigma_{U}^{2})/(Q^{2}+U^{2})$ where $\delta Q$, and $\delta U$ are the uncertainties for Stokes \textit{Q} and \textit{U}.

The polarization position angles $\theta$, measured from north through east on the plane of the sky, were calculated using the relation 
\begin{equation}
{\theta = \frac{1}{2}{\rm tan}^{-1}\frac{U}{Q}} \, ,
\label{eq:theta}
\end{equation}
\noindent
and the corresponding uncertainties in $\theta$ were calculated using

\begin{equation}
   {\delta\theta = \frac{1}{2}\frac{\sqrt{Q^2\delta U^2+ U^2\delta Q^2}}{(Q^2+U^2)}} \,\, .
\label{eq:dtheta}
\end{equation}

The plane-of-sky orientation of the B-field is inferred by rotating the polarization angles by 90$^{\circ}$.
% As mentioned in the Introduction, this rotation is due to the fact that the polarization is caused by elongated thermal dust grains aligned perpendicular to the B-field \citep[see][and references therein]{ALV2015}.

\subsection{Ancilliary Data}
{In our analyses, we have adopted the optical (R-band) polarisation data with a polarisation-to-error ratio of {($P/\delta P$) > 2} from \cite{2013MNRAS.432.1502S} that was observed using the ARIES Imaging Polarimeter mounted at the Cassegrain focus of the 104 cm Sampurnanand telescope, India. We have also used the \textit{Herschel} dust continuum maps to derive the H$_2$ column density and dust temperature maps for this region (see Section \ref{dust_temp} for details).}

\section{Result and Discussion}\label{sec3}

\begin{figure*}
\begin{center}
\includegraphics[width=1\linewidth]{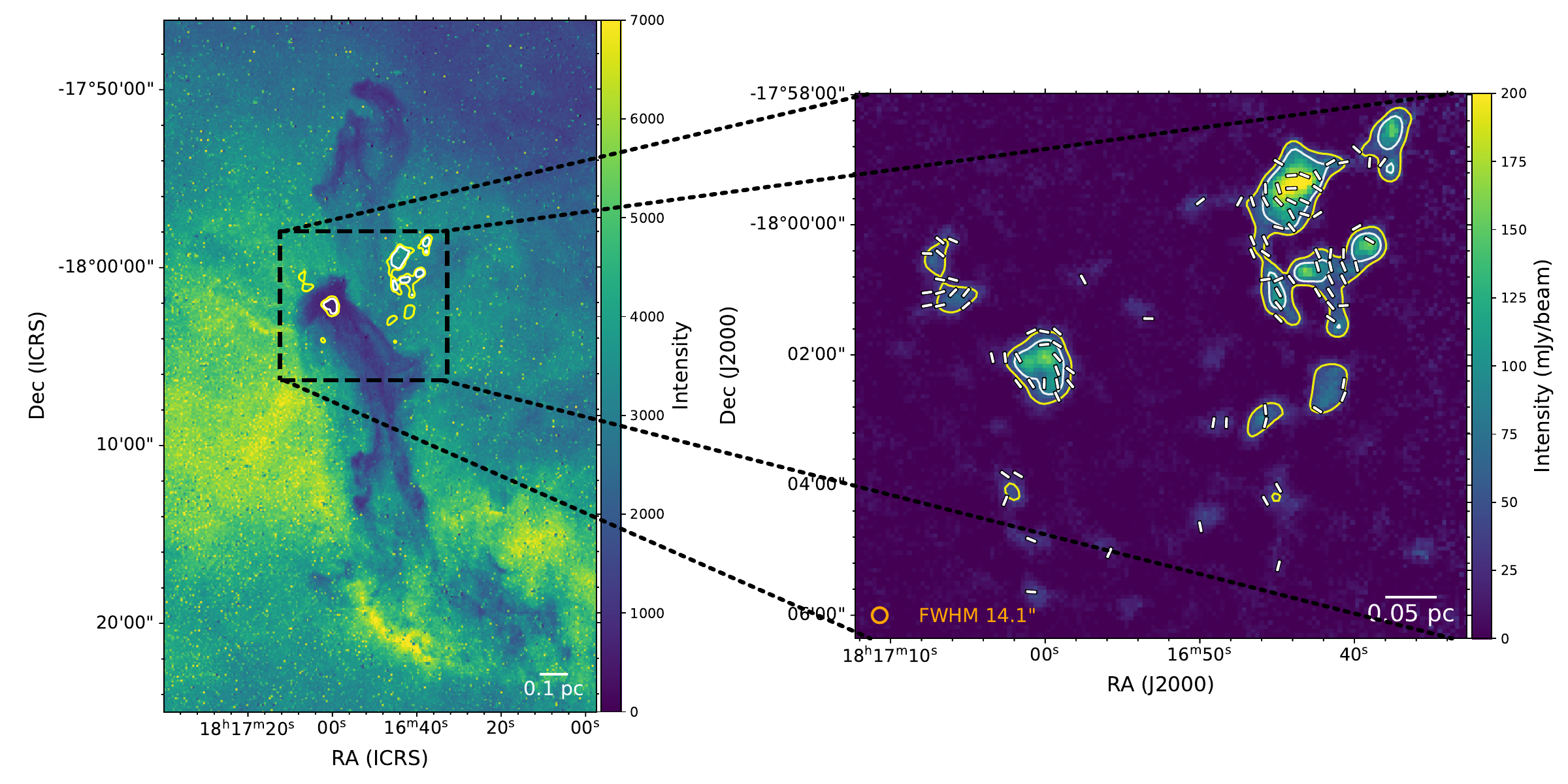}
\caption{Left: DSS red continuum-subtracted H$\alpha$ image of the L328 and L331 regions. Right: 850 $\mu$m dust continuum map overlaid with B-field vectors obtained from SCUBA-2/POL-2, with the orange circle indicating the beam size. The dotted lines outline the zoomed region. Scale bars are shown in the bottom right of both panels. The yellow contour is at 35 Jy/beam, and the white contour is at 70 Jy/beam, taken from the 850 µm continuum image in both images.}\label{Fig:zoom_op_sb1}
\end{center} 
\end{figure*}

\begin{figure*}
\begin{center}
\includegraphics[width=1\linewidth]{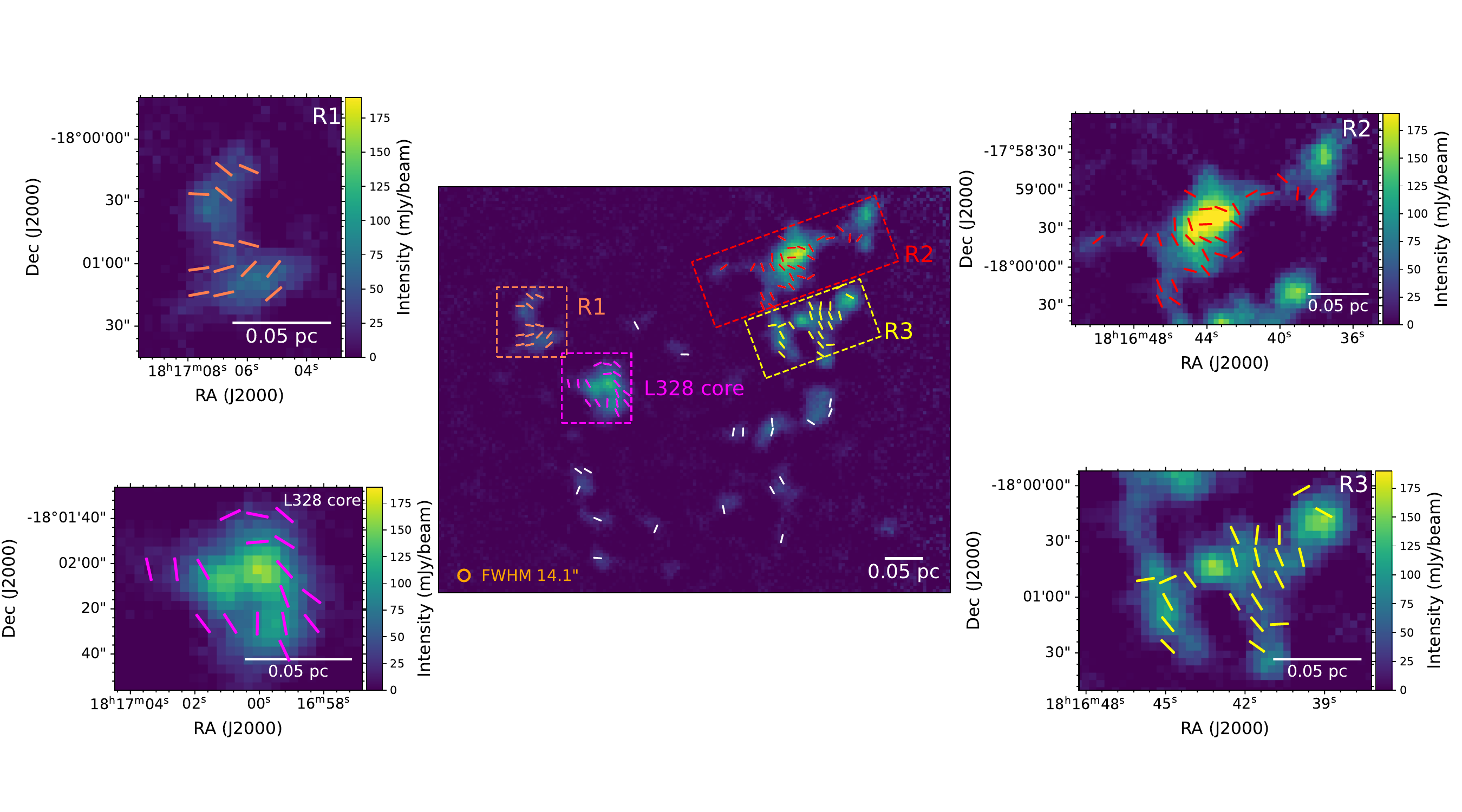}
\caption{The central panel shows the SCUBA-2/POL-2 850 $\mu$m dust continuum map overlaid with B-field vectors, highlighting the regions R1, R2, R3, and the L328 core as introduced in the text. The surrounding panels show the individual zoomed-in images of these regions with B-field vectors overlaid. The scale bar is shown at the bottom left of each image.}\label{Fig:4_region}
\end{center}
\end{figure*}

 In Figure \ref{Fig:zoom_op_sb1}, the left panel shows a DSS red continuum-subtracted H$\alpha$ image containing two dark nebulae, L328 (center shown with black dashed box) and L331 (upper one). The right panel provides a zoomed-in dust continuum map of the black-box region at 850 $\mu$m observations obtained with the JCMT. The white line segments show the B-field morphology obtained with POL-2 observations. This whole region is divided into four sub-regions: R1, R2, R3, and the L328 core. These regions are more distinctly outlined and identified in Figure \ref{Fig:4_region} with zoomed-in B-field structures. Notably, regions R2 and R3 seem to be interconnected by a dusty bridge-like structure. Given their proximity to L328, we assume that regions R1, R2, and R3 are at the same distances of 270 pc as that of L328 core.

% \begin{figure*}
% \begin{center}
% \includegraphics[width=0.9\linewidth]{fellw_clump.pdf}
% \caption{ B-field vectors from SUBA-2/POL-2 850 $\mu$m  are overlaid on the 850 $\mu$m continuum image. Clumps identified using the FellWalker algorithm are shown as ellipses. The clump numbering follows a right-to-left order.}\label{Fig:clump}
% \end{center}
% \end{figure*}

\begin{figure*}
\begin{center}
\includegraphics[width=0.9\linewidth]{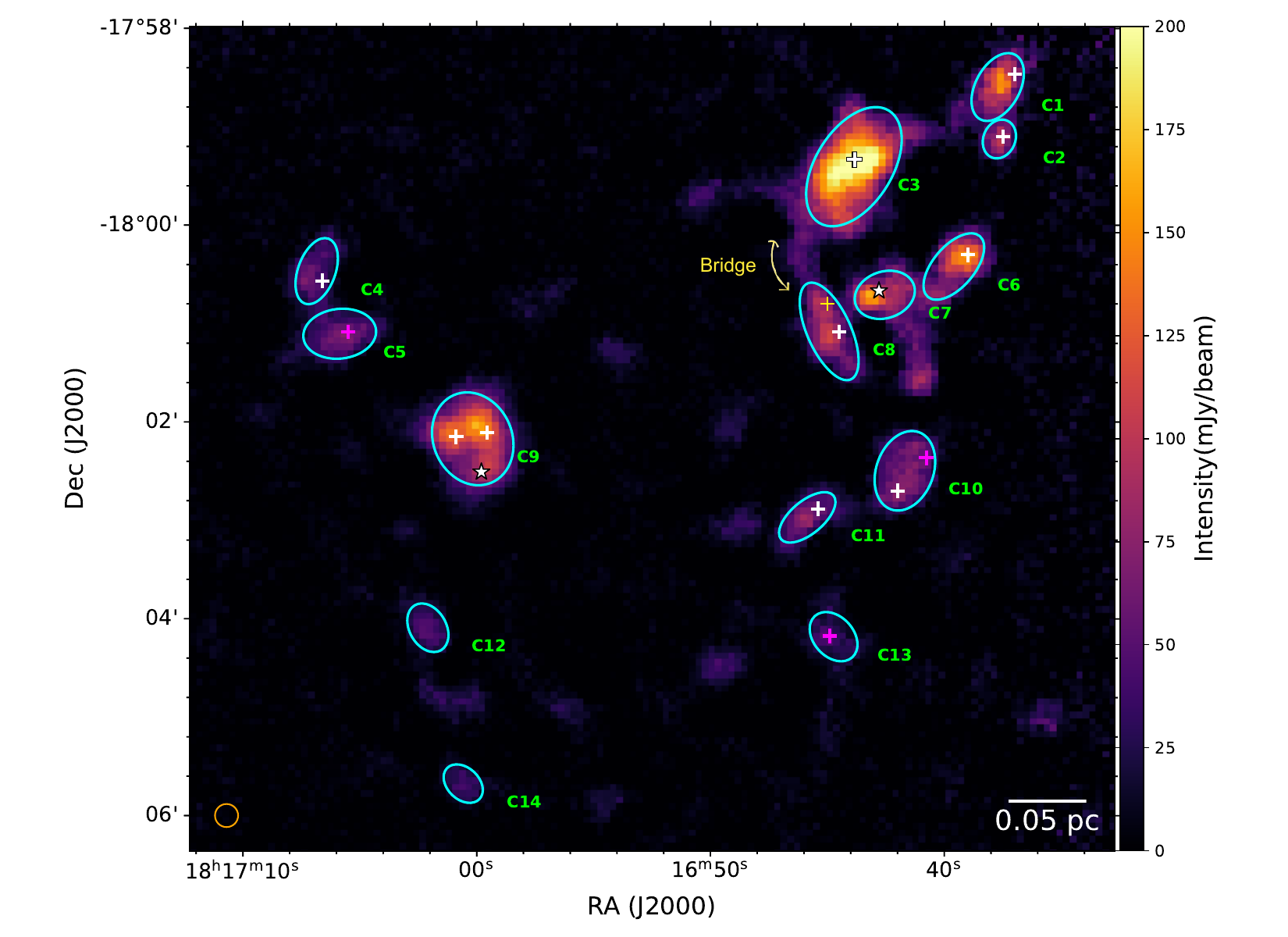}
\caption{Clumps identified using the FellWalker algorithm are shown as ellipses overlaid on the 850 $\mu$m continuum image. The clump numbering follows a right-to-left order. Submillimeter sources are marked with ‘+’, and YSO sources are indicated with ‘*’.} \label{Fig:clump_YSO}
\end{center}
\end{figure*}

\subsection{Dusty structures}
Figure \ref{Fig:clump_YSO} presents the identified clumps overlaid with ellipses, as detected by the FellWalker algorithm.

\subsubsection{Clump finding using FellWalker}

The FellWalker algorithm \citep{2015A&C....10...22B} was used to identify the clumps in the 850 $\mu$m dust continuum image. This algorithm is part of the CUPID package \citep{2007ASPC..376..425B} in the Starlink software\footnote{\href{https://starlink.eao.hawaii.edu/}{https://starlink.eao.hawaii.edu/}}. A total of 14 clumps were identified in the observed region (see Figure \ref{Fig:clump_YSO}). 
In running this algorithm, a source with a peak intensity higher than 3$\sigma_{\rm{rms}}$ and a size larger than the beam size of 14\farcs1 is identified as a real clump. Pixels with
intensities greater than 1.2$\sigma_{\rm{rms}}$ are allowed to be associated with a peak. Two neighboring peaks are considered separate if the dip between the peaks is larger than 1$\sigma_{\rm{rms}}$ deep. The $\sigma_{\rm{rms}}$ value was adopted from \cite{2025MNRAS.539.3493G}. {Ellipses were fitted to each clump based on the 30\% intensity contour level around its peak intensity, and their corresponding major and minor axes (\(\theta_{\text{maj.}}, \theta_{\text{min.}},\) respectively) as well as position angles ($\theta_\mathrm{P.A.}$) were derived to determine their sizes and orientations. $\theta_\mathrm{P.A.}$ is measured from north toward east on the plane of the sky.}

The radius of each detected clump was estimated by considering the deconvolved major and minor axes of the fitted ellipse.  The observed major axis (\(\theta_{\text{maj.}}\)) and minor axis (\(\theta_{\text{min.}}\)) include contributions from the beam size (\(14\farcs1\)). To obtain the intrinsic clump size, we subtract the beam contribution and compute the geometric mean of the deconvolved axes. The final expression for the clump radius is given by:

\begin{figure*}
\begin{center}
\includegraphics[width=1\linewidth]{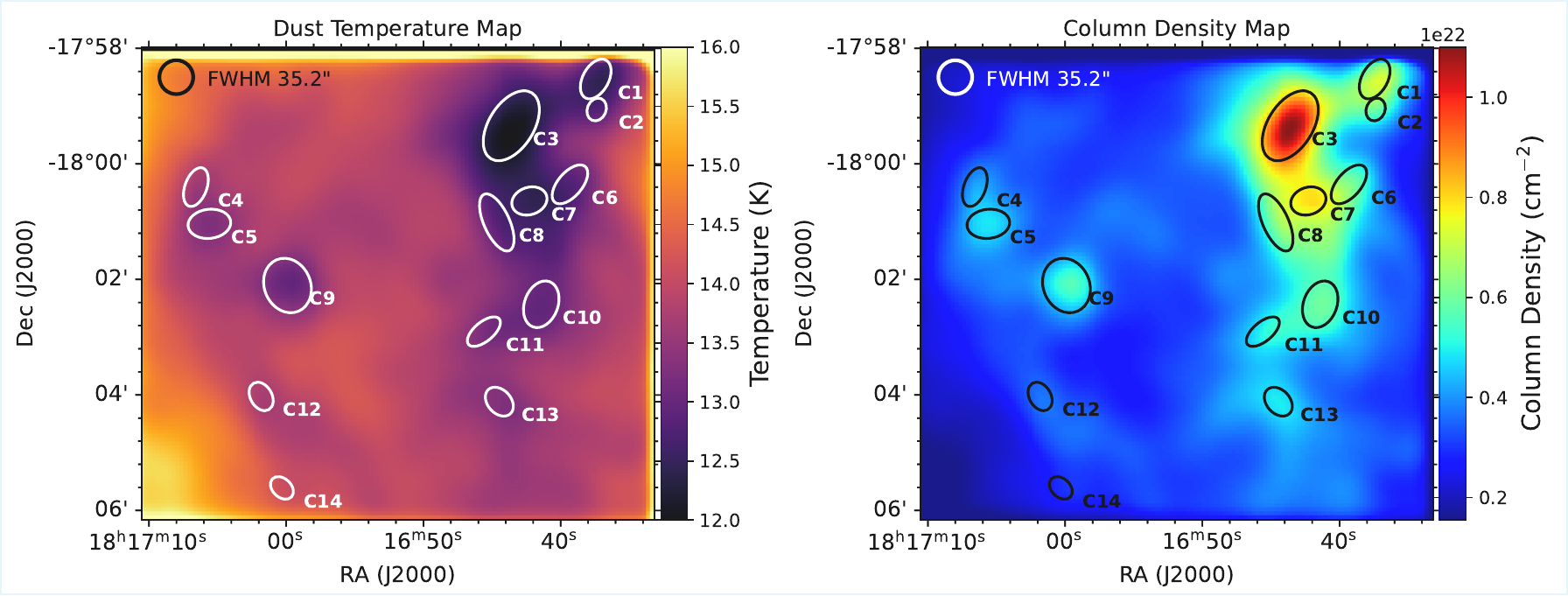}
\caption{The left panel shows the ﬁtted dust temperature $T_{\text{d}}$ map, and the right panel shows the corresponding column density $N({\text{H$_2$}}$) map. The overlaid ellipse on both plots are the clumps from the SCUBA-2 850 $\mu$m dust emission map. Each map is smoothed to the 35\farcs2 beam size of the Herschel 500 $\mu$m observations, which
is shown by the circle in the upper left corner.}\label{Fig:temp_cd_maps_offset}
\end{center}
\end{figure*}

\begin{equation}
R = \sqrt{\frac{\left(\sqrt{\theta_{\text{maj.}}^2 - 14.1^2} \right) \times \left(\sqrt{\theta_{\text{min.}}^2 - 14.1^2} \right)}{4}}.
\end{equation}

Table \ref{tab:clump_parameter} summarizes various physical properties of identified clumps.

\begin{table*}
    \centering
    \scriptsize
    %\footnotesize  %\small   %\tiny
    \begin{tabular}{cccccccccccccc}
    \toprule
    \textbf{Clump } & \textbf{R.A.} & \textbf{Dec.} & $\boldsymbol{\theta_{\rm maj.}}$ & $\boldsymbol{\theta_{\rm min.}}$ & \textbf{Aspect} & \textbf{$\theta_{\rm P.A.}$} & \textbf{Radius} & \textbf{Radius} & \textbf{$I_{\text{peak,850$\boldsymbol{\,\mu}$m}}$ } & \textbf{$T_{\text{dust}}$} & $\boldsymbol{N(\rm{H}_2)}$ & $\boldsymbol{n(\rm{H}_2)}$ & \textbf{Type} \\
    %\footnote{See text for the refrences}} \\

     \textbf{ID}& \textbf{(hh:mm:ss)} & \textbf{(dd:mm:ss)} & \textbf{(\arcsec)} & \textbf{(\arcsec)} & \textbf{Ratio} & \textbf{($\boldsymbol{^{\circ}}$)} & \textbf{(\arcsec)} &  \textbf{(pc)} & \textbf{(mJy/beam)} & \textbf{(K)} & \textbf{($\times$ 10$^{21}$ cm$^{\boldsymbol{-2}}$)}
     &\textbf{($\times$ 10$^{4}$ cm$^{\boldsymbol{-3}}$)}  &\\
    
    \midrule
    C1 & 18:16:37.7  & -17:58:35.8  & 44.75  & 27.01 &{1.66} & 151.39 & 15.64 &0.020 & 152.2 & 12.6$\pm$0.4 & 6.7$\pm$1.2 & 5.43$\pm$0.97 & {smm}  \\
    C2 & 18:16:37.7  & -17:59:07.4  & 24.35  & 19.50 & {1.25} & 157.15 & 8.18 & 0.010 & 102.6 & 12.8$\pm$0.4 & 5.8$\pm$1.0 & 9.40$\pm$1.62 & {smm} \\
    C3 & 18:16:43.9  & -17:59:24.4  & 80.57  & 46.75&{1.72}  & 148.00 & 29.73 & 0.039 & 230.4 & 12.3$\pm$0.4 & 9.2$\pm$1.7 & 3.82$\pm$0.71 & {smm} \\
    C4 & 18:17:06.8  & -18:00:28.1  & 42.15  & 22.79 &{1.85} & 159.94 & 13.33 & 0.018 &68.8 & 13.7$\pm$0.3 & 4.0$\pm$0.6 & 3.60$\pm$0.54 & {smm}  \\
    C5 & 18:17:05.9  & -18:01:06.2  & 44.54  & 30.31 & {1.47}  & 96.29  & 16.83 & 0.022 &72.2 & 13.4$\pm$0.3 & 4.6$\pm$0.7 & 3.39$\pm$0.52 & {smm}  \\
    
    C6 & 18:16:39.6  & -18:00:25.2  & 48.75  & 25.16 & {1.94} & 139.44 & 15.59 & 0.020 &  162.2 & 12.8$\pm$0.4 & 6.3$\pm$1.1 & 5.11$\pm$0.89 & {smm}\\
     % &   &  &   &   &  &  &  & &  &  &  & HIGBAL G013.0145-00.7437 & smm\\
    C7 & 18:16:42.6  & -18:00:42.5  & 37.46  & 28.38 & {1.32} & 108.40 & 14.62 & 0.019 &164.0 & 12.5$\pm$0.4 & 7.7$\pm$1.4& 6.57$\pm$1.19 & {YSO} \\
    C8 & 18:16:44.9  & -18:01:04.8  & 64.20  & 26.61 & {2.41} & 24.35 & 18.80 & 0.025 &113.9 & 13.0$\pm$0.3 & 5.7$\pm$0.9 & 3.69$\pm$0.58 &  {2 smm}  \\
     % &  &  & & &  &  & &  &  &  & & &  & \textbf{SCOPE G013.01-00.77}  \\
    
    C9 & 18:17:00.2  & -18:02:10.3  & 76.2  & 66.6 & {1.15} & 23.30 & 34.91 & 0.036 &167.3 & 13.2$\pm$0.3 & 4.9$\pm$0.7 & 2.41$\pm$0.32 &  {2 smm \&} \\
    &  &  &  &  & &  &  & & &  &  &  & {1 YSO}  \\
    C10 & 18:16:41.7  & -18:02:29.8  & 50.06  & 34.84 & {1.44} & 159.88  & 19.56& 0.025 &77.7 & 13.0$\pm$0.4 & 5.7$\pm$0.9 & 3.50$\pm$0.58 &  {2 smm}  \\
     % & &   &   &  &  & &  & & & & & & & \textbf{HIGALBM G012.9856-00.7651} \\
    C11 & 18:16:45.9  & -18:02:58.2  & 41.58  & 20.02 & {2.08} & 129.79  & 11.79&  0.015 & 87.4 & 13.3$\pm$0.4 & 4.9$\pm$0.7 & 5.29$\pm$0.76 & {smm} \\
    C12 & 18:17:02.1  & -18:04:05.5  & 31.75  & 22.36 & {1.42} & 30.26 & 11.11& 0.015 & 48.9 & 13.8$\pm$0.3 & 3.5$\pm$0.5  & 3.78$\pm$0.54  &-  \\
    C13 & 18:16:44.7  & -18:04:10.9  & 33.97  & 24.39 & {1.39} & 42.32 & 12.40 & 0.016 & 47.7 & 13.3$\pm$0.4 & 4.8$\pm$0.7 & 4.86$\pm$0.71 & {smm}  \\
    C14 & 18:17:00.6  & -18:05:40.6  & 27.33  & 19.38 & {1.41} & 46.54 & 8.82 & 0.012 & 41.5 & 14.2$\pm$0.3 & 2.9$\pm$0.4 &3.92$\pm$0.54 &- \\
    \bottomrule
    \end{tabular}
    \caption{Table of ellipse parameters, including sexagesimal coordinates of centroid positions, major and minor axes, position angles, radius, peak intensity values, dust temperature, column density,  volume density, {classification of source,} and SIMBAD identification (see Section \ref{sec:simbad}) for references.}
    \label{tab:clump_parameter}
\end{table*}

\subsubsection{Matching of Clumps with SIMBAD database} \label{sec:simbad}

{We cross-matched the identified clumps with the help of the SIMBAD database \footnote{\href{https://simbad.cds.unistra.fr/simbad/}{https://simbad.cds.unistra.fr/simbad/}} and found association with the Young Stellar Objects (YSOs) and sub-millimeter (smm) sources (see Figure \ref{Fig:clump_YSO} and Table \ref{tab:clump_parameter}).
 Clumps C1, C2, and C3, are also identified as sub-millimeter (smm) sources: HIGALBM G013.0357-00.7211 \citep{2017MNRAS.471..100E}, SCOPE G013.06-00.73 \citep{2019MNRAS.485.2895E}, and AGAL G013.034-00.749 \citep{2018MNRAS.473.1059U},  respectively.} 
 For AGAL G013.034-00.749, the dust temperature is 12.8 K and the column density $N$(H$_{2}$) is 22.482 (in log scale). 
 %The systematic velocity (V$_{\rm lsr}$) of this source is 44.5 km s$^{-1}$. 
 The line width of C$^{18}$O (2--1) is 0.68 km s$^{-1}$, the line width of $^{13}$CO (2--1) is 1.55 km s$^{-1}$, and the optical depth of 1 \citep{2018MNRAS.473.1059U}.

The clumps C4 and C5, are also identified as smm source named SCOPE G013.06-00.84 \citep{2019MNRAS.485.2895E}, and {HIGBAL G013.0510-00.8413} \citep{2017MNRAS.471..100E}, respectively. %\reference

{Clumps C6, C7, and C8 are associated with SCOPE G013.01-00.74
% , which is also known as JCMTLSP J181638.9-180020
\citep{2019MNRAS.485.2895E}, AGAL G013.011-00.757  which is classified as a YSO \citep{2018MNRAS.473.1059U}, and SCOPE G013.01-00.77 as well as JCMTLSP J181644.9-180049 \citep{2017MNRAS.469.2163E}, respectively.
For AGAL G013.034-00.757, %the distance is 3.02 kpc,
the dust temperature is 13.6 K and the column density 
$N$(H$_{2}$) is 22.355 (in log scale).} 
% The V$_{lsr}$ of this source is 44.0 km/s.
The line width of C$^{18}$O (2--1) is 1.94 km s$^{-1}$, the line width of $^{13}$CO (2--1) is 1.60 km s$^{-1}$, and the optical depth is 0.46 \citep{2018MNRAS.473.1059U}.

{Clump C9 corresponds to the L328 core, which is having S1 (smm), S2 (YSO), and S3 (smm) sub-cores \citep{2007AJ....133.1560W}. The clump C10 has 2 smm sources: SCOPE G012.98-00.77 and HIGALBM G012.9856-00.7651. Clumps C11 and C13 are also identified as smm source as SCOPE G012.99-00.78 and HIGALBM G012.9670-00.7940, respectively.}

\begin{figure*}
\begin{center}
\includegraphics[width=1\linewidth]{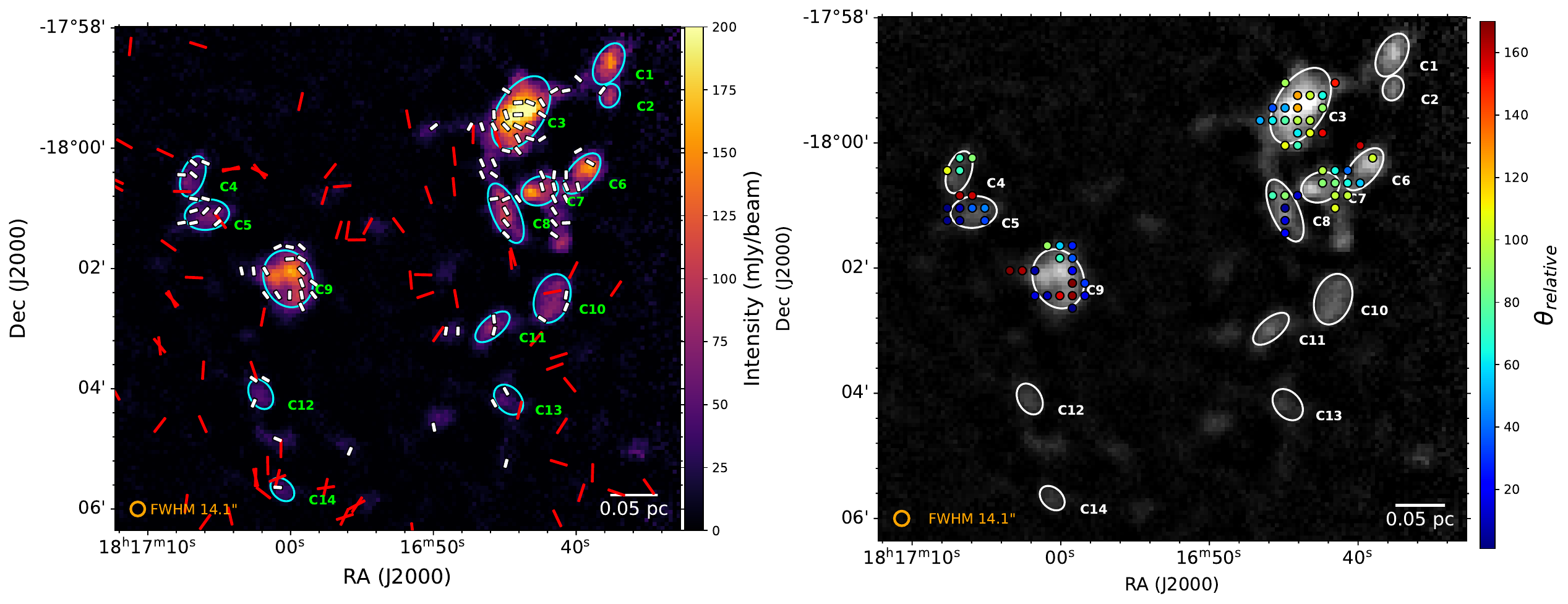}
\caption{Left: 850 $\mu$m dust continuum map overplotted with B-field vectors from SCUBA-2/POL-2 (white) and optical observations (red), with the orange circle indicating the beam size. Right: The SCUBA-2/POL-2 850 $\mu$m dust continuum map is overplotted with color coded circles representing the relative angle between the clump’s major axis and the mean B-field orientation. The color scale spans from 0$^\circ$ (blue) to 180$^\circ$ (red). The orange circle at the bottom left indicates the beam size, and the scale bar at the bottom right.}\label{Fig:Optical_submm_clump}
\end{center}
\end{figure*}

\subsection{Column Densities and Dust Temperatures} \label{dust_temp}
% add offset thing
This region was observed by \(Herschel\) space observatory with Photodetector Array Camera \& Spectrograph (PACS) at 100 and 160 $\mu$m and with Spectral and Photometric Imaging Receiver (SPIRE) at 250, 350, and 500 $\mu$m. We use these archival Herschel data, combined with our JCMT data at 850 $\mu$m, to ﬁt a modiﬁed blackbody function (see Equation  \ref{eq:blackbody}) for the dust emission. The \(Herschel\)/PACS, \(Herschel\)/SPIRE, and JCMT 850 $\mu$m images were smoothed to the SPIRE 500 $\mu$m FWHM beam size of 35\farcs2\,(because it is the maximum beam size) and then re-projected on a common grid (4\arcsec, same as JCMT 850 $\mu$m image). The SPIRE data we obtained from the archive have been zero-point corrected using $Planck$ data while such corrections have not been made for the PACS data. We applied an offset correction to the PACS data by adding the background flux level to restore the missing large-scale emission. The spectral energy distribution (SED) for each pixel was fitted assuming the following formula for a modified blackbody emission \citep[see][]{2008A&A...487..993K}:

\begin{equation} \label{eq:blackbody}
S_\nu = B_\nu(T_\text{d})\left(1 - e^{-\tau_\nu}\right),
\end{equation}
where,
\begin{equation}
 B_\nu(T_d) = \frac{2h\nu^3}{c^2} \frac{1}{e^{\frac{h\nu}{k_BT_d}} - 1},
\end{equation}

\begin{equation}
\tau_\nu = \mu_{\text{H$_2$}} m_\text{H} k_\nu N({\text{H$_2$}}),
\end{equation}
and
\begin{equation} \label{eq:kappa}
k_\nu = k_0 \left( \frac{\nu}{\nu_0} \right)^\beta,
\end{equation}
where S$_\nu$ is the measured flux at the observed frequency $\nu$, $B_\nu$ ($T_\text{d}$) is the Planck function for a dust temperature $T_\text{d}$, $\tau_\nu$ is the optical depth, $\mu_{\mathrm{H_2}}$ is the mean molecular weight of the hydrogen gas in the cloud, $m_{\text{H}}$ is the mass of a hydrogen atom, $N({\text{H$_2$}})$ is the column density, and $\kappa_\nu$ is the dust opacity (absorption coefficient). We use a value of 2.8 for $\mu_{\text{H$_2$}}$, and $\kappa_\nu$ was calculated for each frequency observed using Equation \ref{eq:kappa}, where $\beta$ is the emissivity spectral index of the dust, and we assume $\kappa_0 = 0.1 \,\text{cm}^2\,\text{g}^{-1}$ and $\nu_0 = 10^{12}\,\text{Hz}$ \citep{1990AJ.....99..924B}.

The SED fitting for each pixel was done by fixing $\beta$ as 1.8, while the dust temperature $T_{\rm d}$ and the column density $N({\text{H$_2$}})$ were fitted simultaneously using the \texttt{curve\_fit} function. The resulting temperature and column density maps obtained through this procedure are shown in Figure \ref{Fig:temp_cd_maps_offset}.

\begin{figure*}
\begin{center}
\includegraphics[width=1\linewidth]{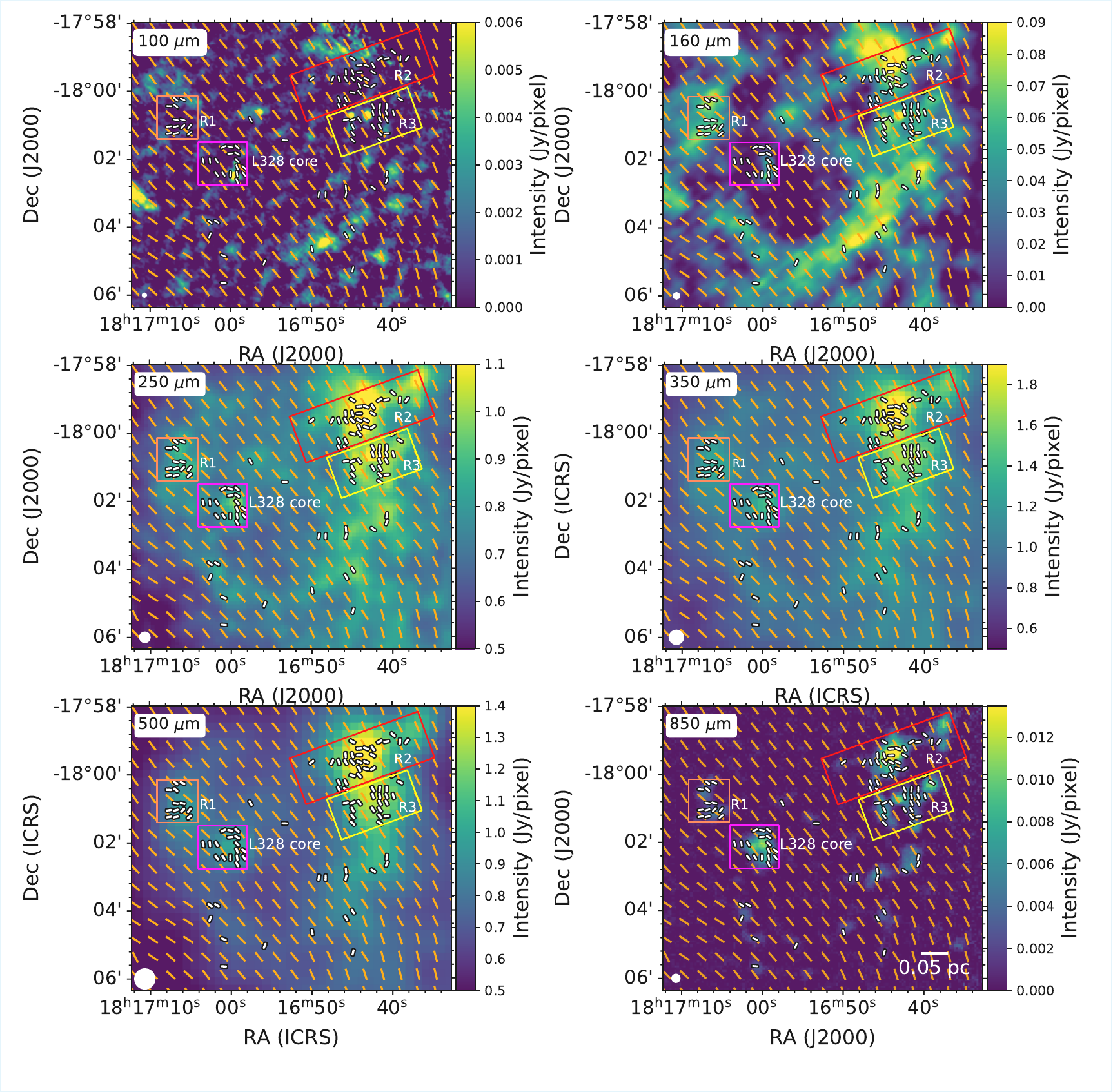}
\caption{Dust emission maps at 100 and 160 $\mu$m from \textit{Herschel}/PACS, and at 250, 350, and 500 $\mu$m from \textit{Herschel}/SPIRE data archives. The 850 $\mu$m map is from JCMT/SCUBA-2 observations in this work. All maps are overplotted with white-colored B-field vectors obtained from the SCUBA-2/POL-2. In addition to this, all maps are also overplotted with orange-colored vectors showing Planck polarization. The white circle in the bottom left of each panel indicates the beam size.}\label{Fig:JCMT_Planck_Jy_pixel_subplot}
\label{}
\end{center}
\end{figure*}

\begin{figure}
\begin{center}
\includegraphics[width=1\linewidth]{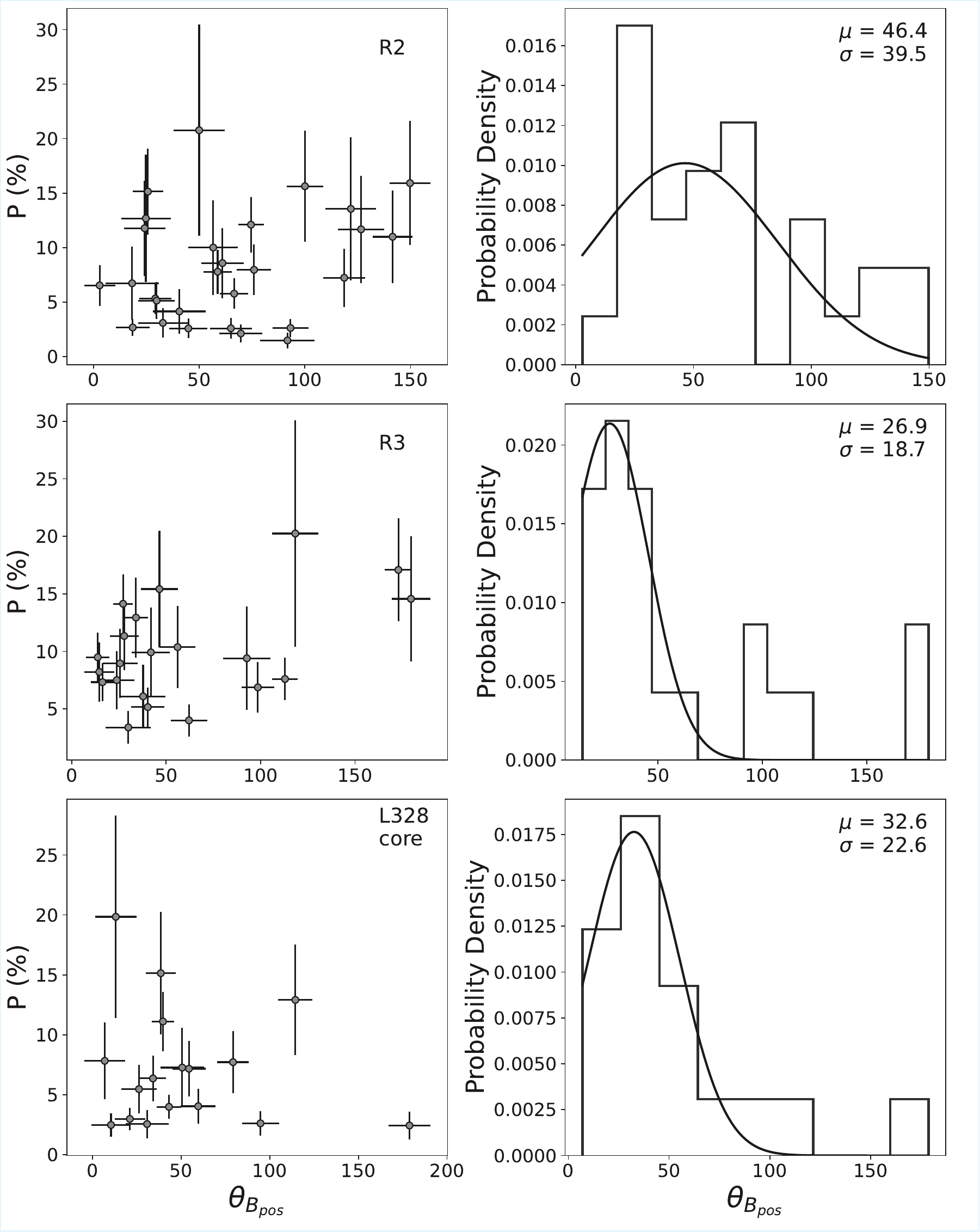}
\caption{The left panel shows the degree of polarization versus the position angle of the B-field for regions R2, R3, and the L328 core. The right panel presents Gaussian-fitted histograms of the B-field position angles with their mean and \textbf{standard deviation.}} \label{Fig:all4}
\end{center}
\end{figure}

The identified clumps are overplotted on the dust temperature map and column density map, as shown in Figure \ref{Fig:temp_cd_maps_offset}. The average dust temperature $T_{\rm{d}}$ and average column density $N({\rm{H_2}}$) for each clump are listed in Table \ref{tab:clump_parameter}. The uncertainties in $T_{\rm{d}}$ and $N({\rm{H_2}})$ are estimated from the diagonal elements of the covariance matrix obtained during the SED fitting. For clump C3, the average $T_{\rm {d}}$ is 12.3 K, which is consistent with the previously determined value of 12.8 K  \citep{2018MNRAS.473.1059U}. The average $N({\rm{H_2}})$ is 9 $\times$ 10$^{21}$ cm$^{-2}$ which is 3 times lower than the value of 30 $\times$ 10$^{21}$ cm$^{-2}$ reported by \citet{2018MNRAS.473.1059U}. For the L328 core, the average  $T_{\rm {d}}$ is 13.2 K, which is fairly consistent with the 11.5 K reported by \cite{2025MNRAS.539.3493G}.
This discrepancy may be attributed to the offset correction applied to the 100 $\mu$m and 160 $\mu$m images prior to the SED fitting.

\subsection{Magnetic Field Morphology}

\subsubsection{B-fields in clumps}

{To evaluate whether the B-fields within the clumps are aligned or perpendicular to their major axis, we compute the relative angle between the position angle of the clump ($\theta_{\text{P.A.}}$) and the  B-field orientation ($\theta_{B_{\text{pos}}}$), defined as:}
\begin{equation}
|\theta_{\text{relative}}| = \left| \theta_{B_{\text{pos}}} - \theta_{\text{P.A.}} \right|
\end{equation}

The right panel of Figure \ref{Fig:Optical_submm_clump}, shows the 850 $\mu$m dust continuum map observed by SCUBA-2/POL-2, illustrating the relative orientation ($\theta_{\text{relative}}$) of the position of B-field ($\theta_{B_{\text{pos}}}$) with respect to the clump ($\theta_{\text{P.A.}}$). The color scale, ranging from 0$^\circ$ to 180$^\circ$, highlights the relative alignment between the B-fields and the clump major axes. The values close to 0$^\circ$ or 180$^\circ$ indicates the parallel alignment, while values near 90$^\circ$ show perpendicular alignment. The parallel configuration suggests that the B-fields may facilitate the channelling of material along the major axis of clump, helping in elongation and subsequent fragmentation \citep{2016A&A...586A.138P}. The perpendicular configuration may suggest that magnetic pressure is playing a role in resisting gravitational collapse, under the assumption that matter is flowing through the major axis of the clump \citep{2019FrASS...6....5H}.

{For clumps C1 and C2, there are no B-field vectors to analyse. Clump C3 exhibits a mix of all colors, from blue to red, showing random nature. This may hint at a dynamically active region.
In clump C4, the dominance of green color suggests a perpendicular B-field orientation relative to the major axis. Conversely, clump C5 shows the dominance of blue color and red color indicating parallel alignment. Clump C6 displays color from blue to red, indicative of a random distribution, which can be due to limited number of B-field vectors. Clump C7 is dominated by green shades, suggesting perpendicular alignment. The presence of a YSO within C7 suggests that gravitational collapse is occurring. Clumps C8 and C9 are dominated by blue and red shades, pointing to a preferential parallel alignment.  These clumps also show evidence of fragmentation, which may be facilitated by flows along the clump axes. Finally, clumps C10 to C14 lack a sufficient number of B-field vectors and were therefore excluded from further analysis.}

{For the statistical analysis, we first computed the mean B-field position angles for each clump, which are derived by fitting a Gaussian to the histograms of position angles of B-field, and their values are listed in the fourth column of Table~\ref{tab:clump_aspect_pa}. From these, we calculated the mean $\theta_{\text{relative}}$, which are presented in the fifth column of Table~\ref{tab:clump_aspect_pa}. The distribution of $\theta_{\rm relative}$ is consistent with the qualitative trends discussed above, confirming the inferred alignment between the B-fields and the clump major axes.}

\begin{table}
    \centering
    \scriptsize
    \begin{tabular}{cccccc}
    \toprule
    \textbf{Clump ID} & \textbf{Aspect Ratio} & \textbf{$\theta_{\rm P.A.}$ ($^\circ$)} & \textbf{$\theta_{B_{\text{pos}}}$ ($^\circ$)} & \textbf{$\theta_{\text{relative}}$ ($^\circ$)} & \textbf{Alignment} \\
    \midrule
    \textbf{C3}  & {1.72} & {148.00} & {51.9 $\pm$ 34.2} & {96.1 $\pm$ 34.2} & {Random} \\
    {C4}  & {1.85} & {159.94} & {58.5 $\pm$ 3.7}  & {101.4 $\pm$ 3.7}  & {$\perp$} \\
    {C5}  & {1.47} & {96.29}  & {113.0 $\pm$ 21.6} & {16.7 $\pm$ 21.6}   & {$\parallel$} \\
    {C6}  & {1.94} & {139.44} & {24.9 $\pm$ 32.3} & {114.5 $\pm$ 32.3} & {Random} \\
    {C7}  & {1.32} & {108.40} & {20.7 $\pm$ 8.9}  & {87.7 $\pm$ 8.9}   & {$\perp$} \\
    {C8}  & {2.41} & {24.35}  & {35.1 $\pm$ 5.7}  & {10.8 $\pm$ 5.7}   & {$\parallel$} \\
    {C9}  & {1.15} & {23.30}  & {32.6 $\pm$ 22.6} & {9.3 $\pm$ 22.6}   & {$\parallel$} \\
    \bottomrule
    \end{tabular}
    \caption{{Aspect ratios, position angles, mean B-field orientations, and relative angles for the clumps. The final column indicates whether the B-fields are preferentially parallel ($\parallel$), perpendicular ($\perp$), or randomly oriented with respect to the clump major axis.}}
    \label{tab:clump_aspect_pa}
\end{table}

\subsubsection{B-fields at different scales}
The left panel of Figure \ref{Fig:Optical_submm_clump} illustrates that the B-fields at parsec scales, as seen in optical measurements (red-colored vectors) and on sub-parsec scales observed in JCMT measurements (white-colored vectors). The red vector appears to show more randomness compared to the white ones. This suggests that B-fields are getting more uniform on sub-parsec scales.

Figure \ref{Fig:JCMT_Planck_Jy_pixel_subplot} presents dust emission maps at different wavelengths obtained from $Herschel$ data archive. At 100 $\mu$m, the emission is dominated by hotter dust. As we move to 160 $\mu$m, 250 $\mu$m, 350 $\mu$m, and 500 $\mu$m, the emission reveals a more colder dust, and we start to see the cloud structures. We noticed that dust structure in general seems to follow a spherical shell morphology. There is a hint of this shell like structure present in 100 and 850 $\mu$m too. 

The large scale B-fields mapped with $Planck$ in orange-colored vectors are plotted on all the subplots. The B-fields on these scales are quite uniform and the B-fields revealed using 850 $\mu$m (in white-colored vectors) are clearly seen to be present in longer wavelengths (i.e. 250 -- 850 $\mu$m) maps. In region R1 and L328 core, the \(Planck\) map shows a magnetic field orientation of approximately 45°, whereas in regions R2 and R3, it tends closer to 30°.  This consistent large-scale orientation reflects \(Planck\)'s sensitivity to grains exposed to the diffuse interstellar radiation field (ISRF), which extends across large scales but is less sensitive to dense regions. In contrast, the white-colored B-field vectors at 850 $\mu$m trace B-fields associated with denser regions that may be influenced by local radiation sources. {Overlaid on each wavelength map, almost all of these vectors are found to be within the spherical shell seen in $Hershel$ emission, demonstrating that the B-field is closely associated with the denser regions.} This multi-wavelength analysis underscores the connection between B-field orientation and dust continuum properties across different environments. However, detailed analysis of the shell structure is outside the scope of this paper.

\subsubsection{B-fields in different regions}
%Figure \ref{Fig:4_region} shows the regions defined in this work. 

In region R1, the B-field has already been defined in terms of two clumps, C4 and C5, in the previous subsection. Region R2 exhibits more chaotic field structure, as we can see in Figure \ref{Fig:Optical_submm_clump}, and statistically confirmed in Figure \ref{Fig:all4} (panel 2b), with a large dispersion ($\sigma$) of 45.6$^{\circ}$. 
Regions R2 and R3 seem to be connected by a bridge-like diffuse continuum structure at 850 $\mu$m and the field lines seem to follow that bridge.

In region R3, we see that B-field lines more or less maintain a consistent orientation. The mean B-field orientation in this region is 28$^{\circ}$, which may be aligned with the large-scale fields ($\sim$30$^{\circ}$). This may suggest a coherent magnetic structure from parsec to sub-parsec scales. The B-field morphology in the L328 core has already been defined in \cite{2025MNRAS.539.3493G}.

The mean polarization $P$(\%) in regions R1, R2, R3, and L328 core is 11\% $\pm$ 4\%, 9\% $\pm$ 3\%, 10\% $\pm$ 3\%, and 7\% $\pm$ 3\%, respectively. 
In Figure \ref{Fig:all4}, we examine the relationship between the degree of polarization $P$(\%) and the B-field orientation represented by $\theta_{{B_{\rm pos}}}$.
Left column shows that region R2 shows most of $P$ values concentrated between position angles 0-100$^{\circ}$. R3 and L328 core shows most of $P$ values concentrated between position angles 0-50$^{\circ}$. This suggests that most of the dust grains are aligned along the dominant component of the magnetic field, which is around 50$^{\circ}$.
The panels in right column show the Gaussian-fitted histogramic representation of the peak B-field orientations in all the regions.

\begin{figure}
\begin{center}
\includegraphics[width=0.96\linewidth]{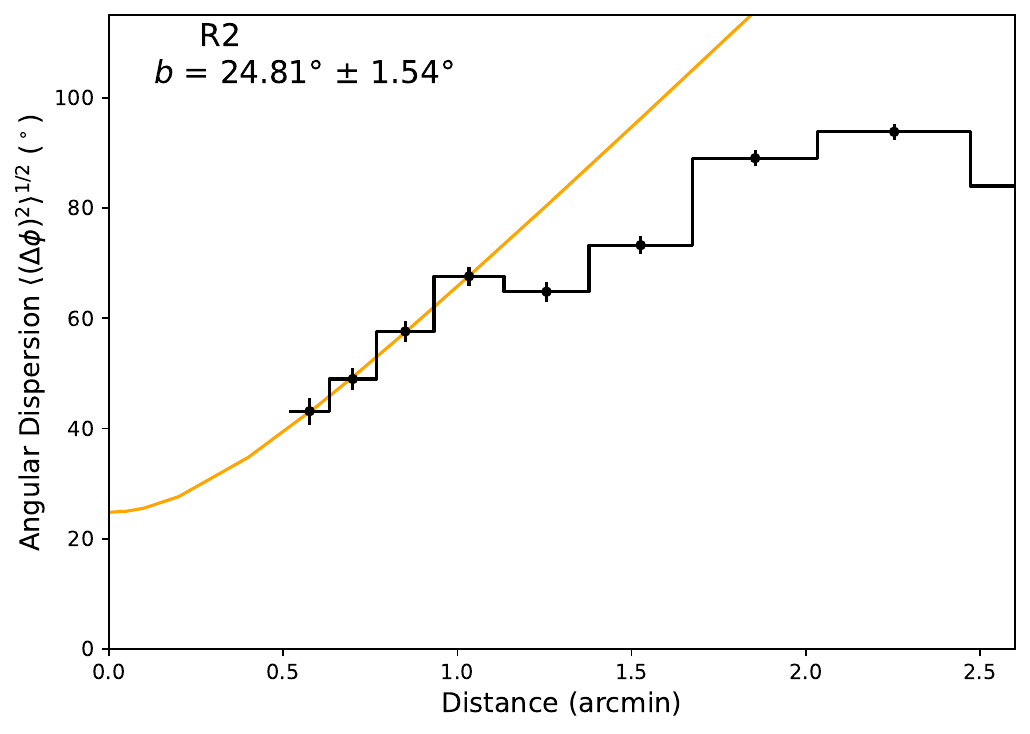}
\caption{Angular dispersion function of R2  region with angle
dispersion segments are shown with black solid circles.
The best ﬁt is shown with an orange line.}\label{Fig:sf_C2}
\end{center}
\end{figure}

\begin{figure*}
\begin{center}
\includegraphics[width=1\linewidth]{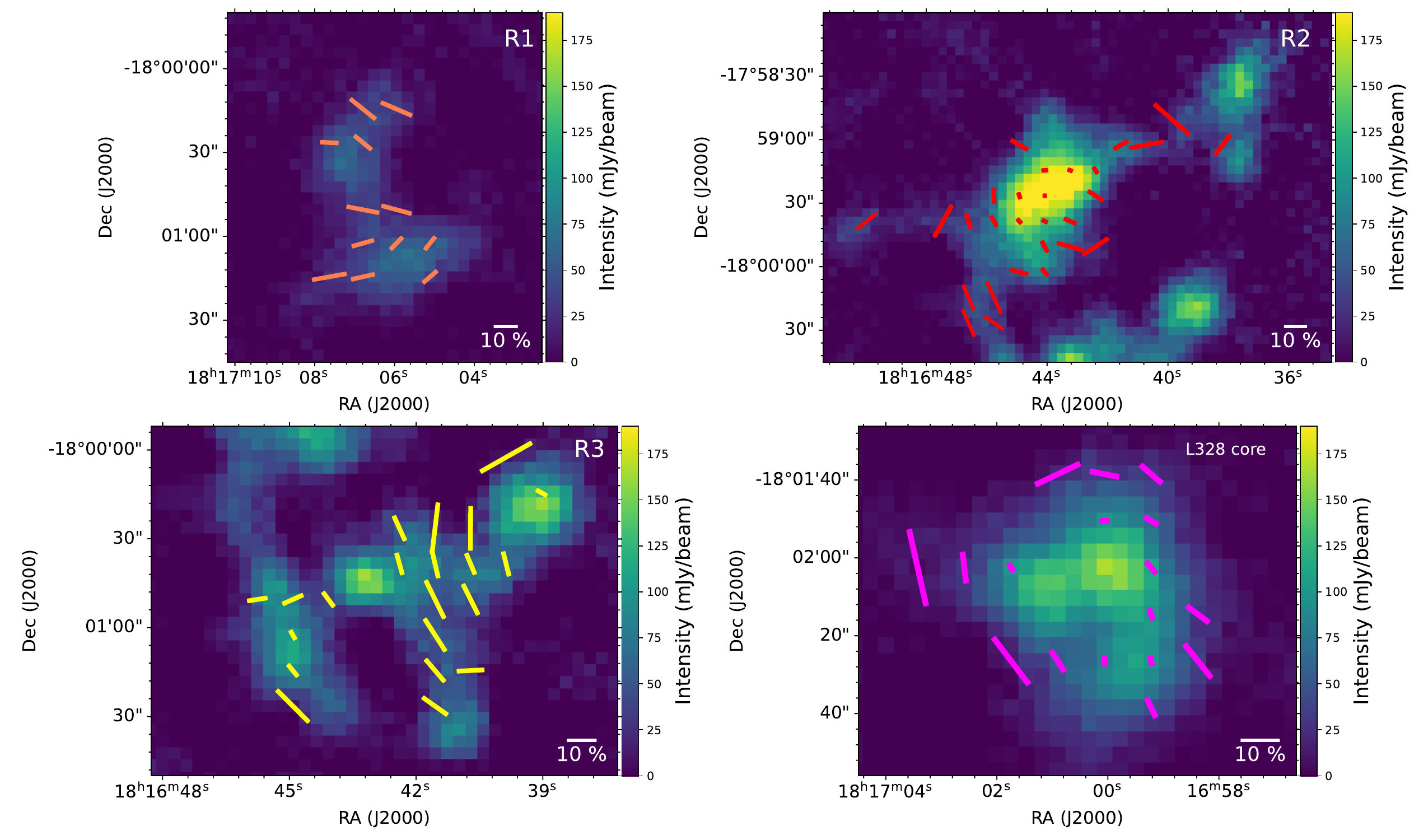}
\caption{All four regions, R1, R2, R3, and L328 core, are shown with polarization vectors, where the length of the vectors indicates the polarization percentage. A polarization scale bar of 10\% is included in each panel for reference.}
\label{Fig:4_region_pol}
\end{center}
\end{figure*}

\subsection{Magnetic field Strength}{\label{mag_strength}}

The plane-of-sky {B-}field strength ($B_{\rm{pos}}$) is estimated using the Davis-Chandrasekhar-Fermi (DCF; \citealt{1951PhRv...81..890D, 1953ApJ...118..113C}) {relation}. The DCF method assumes that small-scale turbulent gas motions distort {B-}field lines, and this distortion is related to the Alfvén Mach number of the gas. Generally, polarization angle dispersion is used to estimate this effect. 
The  B-field strength ($B_0$) is estimated using the classic DCF relation:
\begin{equation}
B_0 = \frac{\sqrt{4\pi\rho} \, \sigma_v}{\delta\theta},
\label{eq:DCF1}
\end{equation}
where $\rho = \mu_{\text{g}} n_{\text{H}_2} m_{\text{H}}$ is the mass density, $\sigma_v$ is the velocity dispersion, and 
$\delta\theta$ is the polarization angle dispersion. However, when $\delta\theta$ exceeds $\sim$25°, the assumption of a weakly perturbed field becomes invalid, potentially leading to underestimated field strengths.
The estimated $B_{\rm pos}$ is thus corrected by a factor $Q$, known as the modified DCF relation \citep{2004ApJ...600..279C}, such that
\begin{equation}
B_{\text{pos}} = Q B_0,
\label{eq:DCF}
\end{equation}
where $Q$ is taken to be 0.5, based on studies using synthetic polarization maps generated from numerically simulated clouds \citep{2001ApJ...546..980}, which suggests that the B-field strength is uncertain by a factor of 2. 

The B-field strength calculation is based on three parameters: volume density $n({\rm{H_2}}$), velocity dispersion   $\sigma_v$(see \ref{velocity}), and polarization angle dispersion  $\delta\theta$ (see \ref{theta}).

\subsubsection{Parameters for B-field Estimation} \label{nh2}
We estimate the volume density of each elliptical clump using the formula, 
\begin{equation}
n({\text{H}_2}) = \frac{N({\text{H}_2})}{2R}\,.
\end{equation}
The average volume density and its uncertainty for each clump are listed in Table \ref{tab:clump_parameter}. The error has been propagated based on the uncertainty in the $N({\rm{H_2}}$). 
For region R2, we compute the volume-weighted average density of clumps C1, C2, and C3, which results in a value of (4.09 $\pm$ 0.60) $\times$ 10$^{4}$ \( \text{cm}^{-3}\). For region R3, the volume-weighted average density of clumps C6, C7, and C8 is (4.71 $\pm$ 0.46) $\times$ 10$^{4}$ \(\text{cm}^{-3}\).

In region R2, C3 is the  largest clump and captures the majority of polarization vectors. Since C3 is associated with AGAL G013.034-00.749, we adopt a C$^{18}$O (2--1) line width of 0.68 km s$^{-1}$. In region R3, we assume the same line width of 0.68 km s$^{-1}$.

%%%%% see at last after checking all sections %%%%%
%\subsubsection{Gas Kinematics} 
%In Region 2, C3 is the  largest clump and captures the majority of polarization vectors. Since, C3 is associated with AGAL G013.034-00.749, we adopt a C$^{18}$O (2-1) line width of 0.68 km/s. In Region 3, we assume the same velocity dispersion of 0.6 km/s. 
\label{velocity}

To estimate the polarization angle dispersion, we fitted a Gaussian function to the histogram of $\theta_{B_{\rm{pos}}}$, as shown in Figure \ref{Fig:all4}. For region R3, the dispersion $\theta_{B_{\rm{pos}}}$ is 18.7$^\circ$, while for region R2, it is  45.6$^\circ$. Since region R2 exhibits a significantly high dispersion, we employed structure function analysis \citep{2009ApJ...696..567H} to determine $\delta\theta$, which provides a better way to quantify the B-field fluctuations.
%%%%%%%%%

\subsubsection{Structure-function analysis} \label{theta}

In the structure function (SF) analysis \citep{2009ApJ...696..567H}, the $B$-field is assumed to consist of a large-scale structured field, $B_0$, and a turbulent component, $\delta B$. The SF method separates out the turbulent component from the large-scale structured field by examining how the dispersion in position angles changes as a function of vector separation $l$. At some scales larger than the turbulent scale $\delta$, $\delta B$ should reach its maximum value. At scales smaller than the scale $d$ (above which $B_0$ varies), the higher-order terms of the Taylor expansion of $B_0$ can be canceled out. When the separation $l$ is $\delta < l < d$, the SF follows the form:
\begin{equation}
 \left <{\Delta \Phi_{\text{tot}}^2} (l)\right> \approx b^2 + m^2 l^2 + \sigma^2_M (l)\,.
\label{eq:sf_formula1}
\end{equation}

In this equation, ${\Delta \Phi_{\text{tot}}^2}(l)$ is the square of the total measured dispersion function, which consists of a constant turbulent contribution ($b^2$), the contribution from the large-scale structured field ($m^2 l^2$), and the contribution of the measured uncertainty in $\theta$ is $\sigma^2_M(l)$. The ratio of the turbulent to the large-scale component of the B-field is given by
\begin{equation}
\frac{\delta B}{B_0} = \left( \frac{b^2}{2-b^2 } \right)^{1/2}\,.
\label{eq:sf_formula}
\end{equation}

Using this ratio, we estimated the B-field strength B$_0$, which is derived from the modified DCF relation incorporating the structure function approach:
\begin{equation}
B_0 = \frac{\sqrt{(2-b^2)\,4\pi\rho} \, \sigma_v}{b}\,.
\label{eq:DCF_sf}
\end{equation} Note that the units of b in equation (\ref{eq:DCF_sf}) are in radians. 
We fitted the quadratic model of the angular dispersion function, as given in equation (\ref{eq:sf_formula1}), to the dispersion of polarization angle differences in region R2 as a function of the distance between pairs of polarization angles, as shown in Figure \ref{Fig:sf_C2}. The fitted b values are found to be 24.81$^\circ$ ± 1.54$^\circ$ (0.43 $\pm$ 0.03 rad) in region R2. In this region, the $B_{\text{pos}}$ is estimated to be 83.4 $\pm$ 10.3 $\mu$G calculated using equation (\ref{eq:DCF_sf}). 

\begin{figure*}
\begin{center}
\includegraphics[width=1\linewidth]{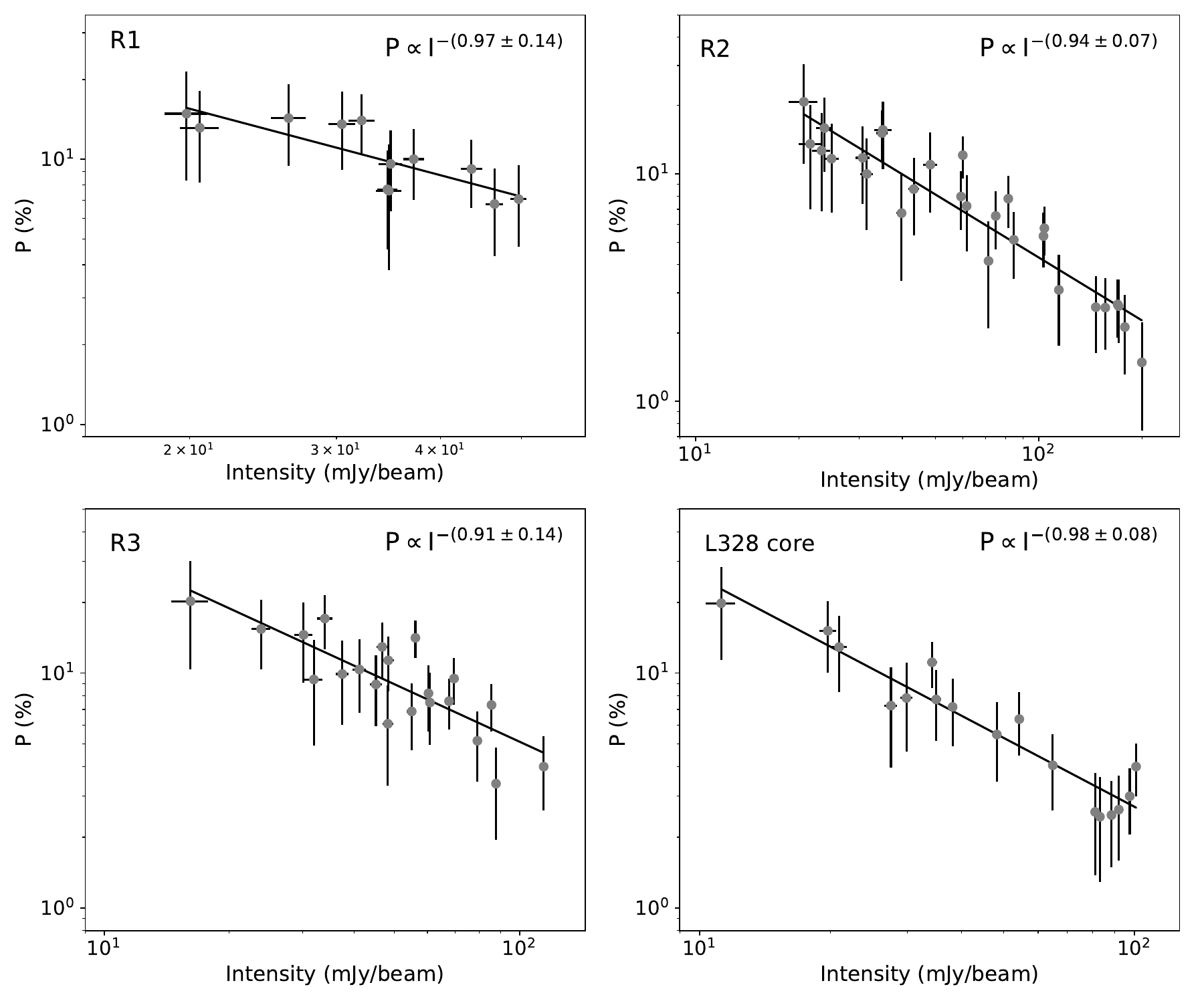}
\caption{ Polarization fraction ($P$) as a function of intensity in the regions R1, R2, R3, and L328 core regions. The data points represent values and uncertainties based on POL-2 measurements. The fitted lines show the relationship $P \propto I^{-\alpha}$, with the corresponding $\alpha$ values indicated in each panel.}\label{Fig:pol_hol}
\end{center}
\end{figure*}

\begin{figure}
\begin{center}
\includegraphics[width=1\linewidth]{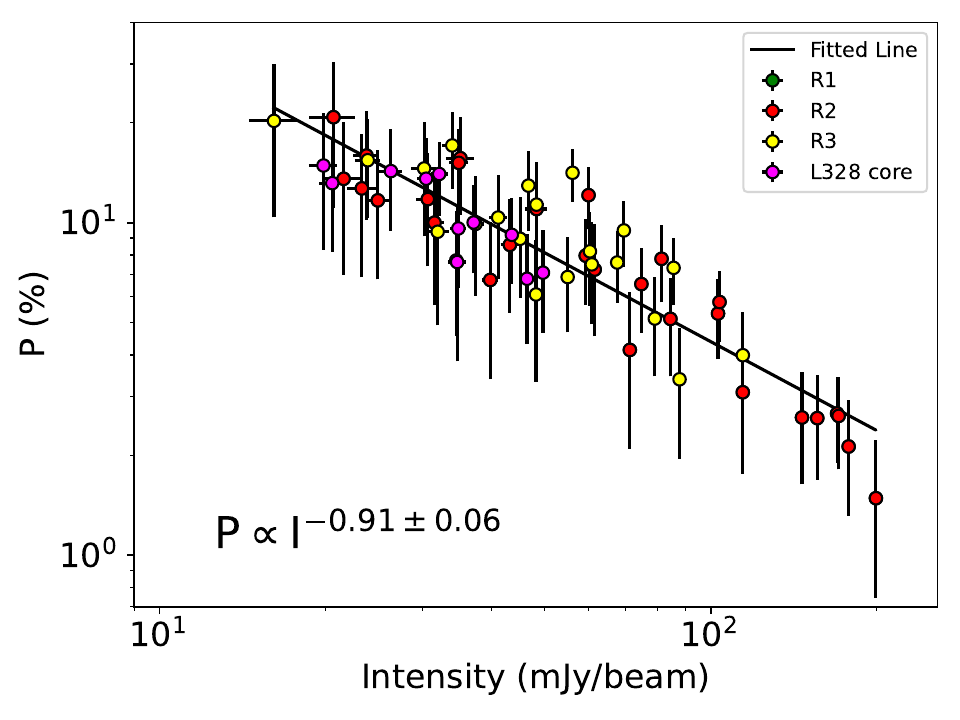}
\caption{Polarization fraction variation with intensity in the regions R1, R2, R3, and L328 core, based on values and uncertainties from POL-2 measurements.}\label{Fig:pol_hol_all_reg}
\end{center}
\end{figure}

{In region R1, the clump C4 and C5 have $B_{\rm pos}$ 143.1 $\pm$ 17.4 $\mu$G and 23.8 $\pm$ 2.9 $\mu$G, respectively. Note that $B_{\rm pos}$ for C4 is quite high compared to the other three regions. This may be attributed to lesser number of B-field vectors within this clumps that leads to smaller dispersion, resulting in potential overestimation of $B_{\text{pos}}$.} For region R3, we determine $\delta\theta$ using the standard deviation of the measured polarization angles, which is 18.7$^{\circ}$ $\pm$ 2.3$^{\circ}$, and the mean of observed errors, 8.8$^{\circ}$.  The calculated $\delta\theta$ is 16.5$^{\circ}$ \citep[see Eq. 8 of][]{2025MNRAS.539.3493G}. The corresponding $B_{\text{pos}}$, calculated from equation (\ref{eq:DCF}), is 83.3 $\pm$ 15.7 $\mu$G.

\subsection{Mass-to-flux Ratio}

% Now that we have B-field strength for regions R2 and R3, we can test the relative importance of B-fields over gravity by calculating the mass-to-flux ratio, which is represented by a parameter $\lambda$ \citep{2004Ap&SS.292..225C}. The parameter $\lambda$ serves as an indicator of the balance between magnetic support and gravitational collapse in a given structure. When $\lambda$ < 1, the structure is considered `magnetically subcritical’, meaning it is supported against gravitational collapse by B-fields. Conversely, when $\lambda$ > 1, the B-field is insufficient to prevent gravitational collapse, and the structure is termed `magnetically supercritical’.

% The value of $\lambda$ is estimated using the the relation given by \citep{2004Ap&SS.292..225C}
% \begin{equation}
% \lambda = 7.6\times10^{-21} \frac{{N({\rm H_{2}})}/{{\rm cm}}^{-2}}{{B_{\rm pos}}/\mu {{\rm G}}}\,\,.
% \end{equation}

Now that we have B-field strength for regions R2 and R3, we can test the relative importance of B-fields over gravity by calculating the mass-to-flux ratio ($\lambda$), as defined by \cite{2004Ap&SS.292..225C}:
\begin{equation}
\lambda = 7.6\times10^{-21} \frac{{N({\rm H_{2}})}/{{\rm cm}}^{-2}}{{B_{\rm pos}}/\mu {{\rm G}}}\,\,.
\end{equation}
The parameter $\lambda$ quantifies the balance between magnetic support and gravitational collapse within a structure. When $\lambda$ < 1, the structure is considered to be `magnetically subcritical’, suggesting that the structure is supported against gravitational collapse by B-fields. In contrast, when $\lambda$ > 1, the B-field is insufficient to prevent gravitational collapse, thus making the structure `magnetically supercritical’.

The average column density $N\rm(H_{2})$ and its uncertainty for each clump are listed in Table \ref{tab:clump_parameter}. The error has been propagated based on the uncertainty in the $N\rm(H_{2})$. {For region R1, the value of $\lambda$ for clump C4 and C5 are 0.2 $\pm$ 0.04 and 1.5 $\pm$ 0.3, respectively.} For region R2, we compute the column-weighted average density of clumps C1, C2, and C3, which results in a value of (8.54 $\pm$ 1.30) $\times$ 10$^{21}$ cm$^{-2}$. For region R3, the (6.39 $\pm$ 0.63) $\times$ 10$^{21}$ cm$^{-2}$. The value of $\lambda$ for regions R2 and R3 comes out to be 0.78 $\pm$ 0.15 and 0.58 $\pm$ 0.12, respectively, suggesting both the regions are magnetically subcritical. The L328 core is found to be magnetically transcritical with a $\lambda$ value of 1.1 $\pm$ 0.2 \citep{2025MNRAS.539.3493G}. 

\subsection{Alfv\'en Mach Number}
The Alfv\'en Mach number ($\mathcal{M}_{\rm A}$) quantifies the relative significance of nonthermal motions with respect to the magnetic field. It is defined as the ratio of the nonthermal (turbulent) velocity dispersion ($\sigma_{\rm NT}$) to the Alfv\'en velocity ($V_{\rm A}$) as defined by \cite{crutcher1999detection}:
\begin{equation}
\mathcal{M}_{\rm A} = \frac{\sigma_{\rm NT}}{V_{\rm A}},
\end{equation}
where $V_{\rm A}$ is expressed as
\begin{equation}
V_{\rm A} = \frac{B_{\rm pos}}{\sqrt{4\pi\rho}}.
\end{equation}

In region R1, the clumps C4 and C5 exhibit \(\mathcal{M}_{\rm A}\) of \(0.26 \pm 0.04\) and \(1.51 \pm 0.22\), respectively.  For regions R2 and R3, the corresponding \(\mathcal{M}_{\rm A}\) values are \(0.54 \pm 0.08\) and \(0.58 \pm 0.11\), while the L328 region shows \(\mathcal{M}_{\rm A} = 0.71 \pm 0.14\). Values of \(\mathcal{M}_{\rm A} < 1\) correspond to sub-Alfvénic regimes, where magnetic fields dominate; \(\mathcal{M}_{\rm A} \approx 1\) indicates a trans-Alfvénic state, where turbulence and magnetic forces are comparable; and \(\mathcal{M}_{\rm A} > 1\) suggests a super-Alfvénic (turbulence-dominated) environment.

Accordingly, clump C4 represents a strongly magnetically dominated region (\(\mathcal{M}_{\rm A} = 0.26\)), whereas C5 (\(\mathcal{M}_{\rm A} = 1.51\)) is turbulence-dominated. However, these values could partly result from uncertainties in \(B_{\rm pos}\) due to the limited number of polarization vectors within clump.  
Regions R2, R3, and L328, with \(\mathcal{M}_{\rm A}\) values 0.5--0.7 lie in the sub-Alfvénic regime, implying that magnetic fields dominate over turbulence. Overall, the predominantly sub-Alfvénic nature of these regions (except C5 clump) indicates that magnetic fields dominate over turbulence, thereby delaying the onset of star formation.

% \textbf{R3, = 0.5$\pm$0.1 km/s,  M = 0.58$\pm$0.11
% R2, = 0.54$\pm$0.08 km/s,     M = 0.54$\pm$0.08
% C4 =  0.98$\pm$0.14 km/s,     M= 0.26$\pm$0.04 
% C5 = 0.17$\pm$0.02 km/s,      M= 1.51$\pm$0.22
% l328= 0.3$\pm$0.06 km/s,      M= 0.71$\pm$0.14}

% add a point in summary and abstract.

\subsection{Polarization as a function of intensities}

Figure \ref{Fig:4_region_pol} shows the four regions with overplotted polarization vectors of varying lengths. We observed that the length of the polarization vectors decreases as we move from lower-intensity to higher-intensity parts in all the regions. Since the length of the vectors corresponds to the polarization fraction, and intensity has a one-to-one correspondence with density, this indicates a drop in polarization fraction toward higher-density regimes. This effect is consistently observed across all four regions, as seen in the figure.

This phenomenon, also known as ``depolarization" or ``polarization hole," can be quantitatively analyzed by comparing the degree of polarization ($P$) with total intensity ($I$) using the relation $P \propto I^{-\alpha}$, as shown in Figure \ref{Fig:pol_hol}. All four panels in the figure demonstrate a negative correlation between $P$ and $I$, with slopes $\alpha$ = 0.97 $\pm$ 0.14, 0.94 $\pm$ 0.07, 0.91 $\pm$ 0.14, and 0.98 $\pm$ 0.08 for the regions R1, R2, R3, and the L328 core, respectively.

Figure \ref{Fig:pol_hol_all_reg} shows the polarization fraction (\(P\)) plotted against intensity (\(I\)) for all four regions (R1, R2, R3, and L328) combined. A clear negative correlation between \(P\) and \(I\) is observed, which follows a power-law relation \(P \propto I^{-\alpha}\), with a slope of \(\alpha = 0.91 \pm 0.06\). This consistent trend across all regions suggests a similar behavior of polarization fraction as a function of intensity, irrespective of local variations in physical conditions such as density, B-fields, and turbulence. The combined data improve the robustness of the statistical fit, demonstrating the universality of the relationship between \(P\) and \(I\) across these regions.

There are several accepted explanations for the observed ``depolarization" effect in starless and star-forming cores. One possible reason is the change in the B-field orientation in denser regions. In denser regions, randomization of gas due to gas-grain collisions can disrupt the alignment of dust grains, reducing alignment efficiency and leading to lower polarization \citep{2007JQSRT.106..225L, 2021ApJ...907...93S}. Another possible explanation is magnetic reconnection, where B-field lines break and reconnect, disrupting the uniformity of the field. It can be because of magnetic tangling also. Since dust grains align with the local B-field, this disruption can reduce their alignment efficiency \citep{1999ApJ...517..700L}.  Additionally, insufficient radiation causing weak RAT \citep{2007MNRAS.378..910L} is another explanation for depolarization.

\section{Summary}\label{summary}

The paper presents an observational study of the L328 core and its nearby clumps by using 850 $\mu$m JCMT/POL-2 dust polarization data. The L328 core and its nearby clumps are divided into four regions for further analysis. The key findings are summarized as follows:

\begin{itemize}
    \item We have identified 14 clumps using the FellWalker algorithm and derived their centroid position, position angle, and radius. 
    \item We estimated the gas column density and the dust temperature of the mapped region by supplementing our 850 $\mu$m data with \textit{Herschel} SPIRE/PACS continuum observations. The average column density and temperature values for each clump are also derived. 
    \item The average volume density for each clump and region was computed. The values for region R2 and R3 are $\sim$ 4.09 $\times$ 10$^{4}$ cm$^{-3}$ and 4.71 $\times$ 10$^{4}$ cm$^{-3}$, respectively.
    \item The B-field morphology is perpendicular for clump C4, parallel for C5 and C8, and chaotic for C3.
    \item In region R3, the B-field morphology at 850 $\mu$m may be linked to the large-scale field structure, as they appear to follow the same direction. 
    \item The magnetic field strength in region R2 is estimated to be 83.4 $\pm$ 10.3 $\mu$G using the structure-function analysis on the modified DCF method. For region R3, the field strength is 83.3 $\pm$ 15.7 $\mu$G, estimated using only the modified DCF method.
    \item The region R2 and R3 are magnetically sub-critical with the $\lambda$ value of 0.78 $\pm$ 0.15 and 0.58 $\pm$ 0.12, respectively, whereas L328 core is transcritical with the value of 1.1 $\pm$ 0.2.
    \item All the regions are found to be sub-Alfvénic except C5 clump.

    % , suggesting both the regions are magnetically subcritical. The L328 core is found to be magnetically transcritical with a $\lambda$ value of 
    % \item A positive correlation is seen between polarization P(\%) and local dispersion $\Delta$B$_{\rm{pos}}$ in regions R2 and R3, which shows dust polarization efficiency is increasing.  In contrast, regions R1 and the L328 core exhibit a negative correlation.                              
    \item The polarization fraction as a function of total intensity is found to be decreasing in the high-density region, indicating depolarization in all the regions with a power-law slope of -0.97, -0.94, -0.91, and -0.98, respectively. Finally, if we combine all the regions, we still see a slope of -0.91.
\end{itemize}
% Gas kinematics, mention the distance, 
\section*{Acknowledgements}
This research has made use of the SIMBAD database, operated at CDS, Strasbourg, France. We also acknowledge the use of NASA’s SkyView facility {(\href{http://skyview.gsfc.nasa.gov}{http://skyview.gsfc.nasa.gov})} located at NASA Goddard Space Flight Center. J.K. is supported by the Royal Society under grant number RF\textbackslash ERE\textbackslash231132, as part of project URF\textbackslash R1\textbackslash211322. C.W.L. was supported by Basic Science Research Program through the National Research Foundation of Korea (NRF) funded by the Ministry of Education, Science, and Technology.
The JCMT is operated by the East Asian Observatory on behalf of National Astronomical Observatory of Japan; Academia Sinica Institute of Astronomy and Astrophysics; the Korea Astronomy and Space Science Institute; the Operation, Maintenance and Upgrading Fund for Astronomical Telescopes and Facility Instruments, budgeted from the Ministry of Finance of China. SCUBA-2 and POL-2 were built through grants from the Canada Foundation for Innovation. This research used the facilities of the Canadian Astronomy Data Centre operated by the National Research Council of Canada with the support of the Canadian Space Agency. SPIRE has been developed by a consortium of insti-
tutes led by Cardiff Univ. (UK) and including: Univ. Lethbridge (Canada);
NAOC (China); CEA, LAM (France); IFSI, Univ. Padua (Italy); IAC (Spain);
Stockholm Observatory (Sweden); Imperial College London, RAL, UCL-
MSSL, UKATC, Univ. Sussex (UK); and Caltech, JPL, NHSC, Univ. Colorado
(USA). This development has been supported by national funding agencies:
CSA (Canada); NAOC (China); CEA, CNES, CNRS (France); ASI (Italy);
MCINN (Spain); SNSB (Sweden); STFC, UKSA (UK); and NASA (USA).
PACS has been developed by a consortium of institutes led by MPE (Germany)
and including UVIE (Austria); KUL, CSL, IMEC (Belgium); CEA, OAMP
(France); MPIA (Germany); IFSI, OAP/AOT, OAA/CAISMI, LENS, SISSA
(Italy); IAC (Spain). This development has been supported by the funding
agencies BMVIT (Austria), ESA-PRODEX (Belgium), CEA/CNES (France),
DLR (Germany), ASI (Italy), and CICT/MCT (Spain). 

%%%%%%%%%%%%%%%%%%%%%%%%%%%%%%%%%%%%%%%%%%%%%%%%%%
\section*{Data Availability}
The data used in this article will be shared on reasonable request to the corresponding author.

$Software$: Starlink \citep{currie2014}, SMURF \citep{chapin2013}, APLpy  \citep{aplpy2012,aplpy2019}, Astropy \citep{astropy:2013,astropy:2018,astropy:2022}, SciPy \citep{2020SciPy-NMeth}, PyAstronomy \citep{pya}, Numpy \citep{ harris2020array}.

%%%%%%%%%%%%%%%%%%%% REFERENCES %%%%%%%%%%%%%%%%%%

% The best way to enter references is to use BibTeX:

\bibliographystyle{mnras}
\bibliography{refL328} % if your bibtex file is called example.bib

\bsp	% typesetting comment
\label{lastpage}
\end{document}